# A search on the Nikiforov-Uvarov formalism

**B. Gönül and K. Köksal**

Department of Engineering Physics, University of Gaziantep, 27310, Gaziantep-Türkiye

**Abstract**

An alternative treatment is proposed for the calculations carried out within the frame of Nikiforov-Uvarov method, which removes a drawback in the original theory and by pass some difficulties in solving the Schrödinger equation. The present procedure is illustrated with the example of orthogonal polynomials. The relativistic extension of the formalism is discussed.

## 1. Introduction

Recently there has been renewed interest in solving simple quantum mechanical systems within the framework of the Nikiforov-Uvarov (N-U) method [1]. This algebraic technique is based on solving the second-order linear differential equations, which has been used successfully to solve Schrödinger, Dirac, Klein–Gordon and Duffin–Kemmer–Petiau wave equations in the presence of some well-known central and non-central potentials, for more recent review see [2-11].

The motivation behind the work in this article is the hidden analogy between the mathematical representation of the N-U method and that of the formalism introduced by Levai [12], and is also the recent discussion [13] regarding the equivalence between the two alternative treatments of the Schrödinger equation in the works [12] and [13, 14, and the related references therein] which is the backbone of the scheme introduced here.

The arrangement of this Letter is as follows. In the next section, a brief survey of the original treatment is given within the framework of the N-U model and the related formulae necessary for subsequent sections are collected. In the light of this summarized work, we turn to the presentation of a simple but more effective algorithm for obtaining solutions of solvable potentials. Section 3 involves some applications of the new formalism revealing the success and elegancy of the refined model. To our knowledge, a requirement for such a refinement in order to improve the N-U model calculations leading to a more applicable and powerful calculation scheme have been escaped notice in other publications. In section 3, we also discuss a possible extension of the new procedure to the relativistic domain with the consideration the Klein-Gordon equation. Finally, the results are summarized in the concluding section.

## 2. Formalism

To proceed, we first give a brief summary of the usual N-U model which in general reduces the second-order differential equations to the hypergeometric type with an appropriate coordinate transformation $s = s(r)$

$$F''(s) + \frac{\tilde{\tau}(s)}{\sigma(s)} F'(s) + \frac{\tilde{\sigma}(s)}{\sigma^2(s)} F(s) = 0 \quad , \tag{1}$$

where $\sigma$ and $\tilde{\sigma}$ are at most second degree polynomials, and $\tilde{\tau}$ is a first-degree polynomial. To find the particular solution of (1), one can use the following transformation $F(s) = \phi(s) y(s)$ leading to a hypergeometric type equation

$$\sigma(s) y''(s) + \tau(s) y'(s) + \Lambda y(s) = 0 \quad , \tag{2}$$

where $y(s)$ satisfies the Rodrigues relation

$$y_n(s) = \frac{B_n}{\rho(s)} \frac{d^n}{ds^n} \left[ \sigma^n(s) \rho(s) \right] \quad .$$

In the above equation, $B_n$ is the normalization constant and $\rho$ is the weight function satisfying the condition $(\sigma \rho)' = \tau \rho$. In addition,

$$\frac{\phi'(s)}{\phi(s)} = \frac{\pi(s)}{\sigma(s)} \quad , \quad \tau(s) = \tilde{\tau}(s) + 2\pi(s) \quad , \quad \Lambda_n = -n\tau'(s) - \frac{n(n-1)}{2} \sigma''(s) \quad , \quad n = 0,1,2,.... \tag{3}$$

in which

$$\pi(s) = \frac{\sigma'(s) - \tilde{\tau}(s)}{2} \pm \sqrt{\left( \frac{\sigma'(s) - \tilde{\tau}(s)}{2} \right)^2 - \tilde{\sigma}(s) + k\sigma(s)} \quad , \quad k = \Lambda - \pi'(s) \quad . \tag{4}$$

Here, $\pi$ obviously is a polynomial depending on the transformation function $s(r)$. The determination of $k$ is the essential point in the calculation of $\pi$, for which the discriminant of the square root in (4) is set to zero.

Many of the N-U model calculations, except a few ones such as [3], should deal with a comparison of the two independent $\Lambda$ − values in (3) and (4) to extract the energy spectrum for a potential of interest. The most significant point at this stage is the derivative of $\tau$ in (3) which must be negative to reproduce physically acceptable positive $\Lambda_n$ – values. This compulsory choice restricts the application of the model to the Schrödinger equation for some special functions, which means that the N-U formalism does not work efficiently for all exactly solvable potentials. However, this deficiency disappears supontaneously within the frame of the present method. Moreover, the novel approach introduced seems more simple, flexible and elegant when compared to the original theory, due to the use of a different treatment for the calculation of energy values. Finally, the present scenario suggests a simple algorithm for the definition of corresponding wave functions, unlike the cumbersome relations used in the standard N-U model.

Let us now focus on the introduction of the new procedure. It is well known that many of the special functions of mathematics represent solutions to differential equations of the form in Eq. (1) where the functions $\tilde{\tau}/\sigma$ and $\tilde{\sigma}/\sigma^2$ are well defined for any particular function [15]. Within this context, and bearing in mind the algorithm of the standard N-U formalism, we proceed first with the transformation of the Schrödinger equation to the one like Eq. (1). However, it is stressed that here a more general transformation scheme will be considered, unlike the usual treatment in the N-U model where a specific transformation function $s(r)$ is chosen to solve a particular potential.

Starting with the consideration of the Schrödinger equation $(\hbar = 2m = 1)$,

$$\frac{\Psi''(r)}{\Psi(r)} = V(r) - E \ , \tag{5}$$

and remembering that its solutions generally take the form

$$\Psi(r) = f(r) F[s(r)], \tag{6}$$

which enables us to use a similar mapping as in the well known N-U technique, we show that the substitution of (6) into (5) leads to the second-order differential equation

$$\left( \frac{f''}{f} + \frac{F'' s'^2}{F} + \frac{s'' F'}{F} + 2\frac{F' s' f'}{F f} \right) = V - E \ , \tag{7}$$

that is reduced to the form of Eq.(1)

$$F'' + \left( \frac{s''}{s'^2} + 2\frac{f'}{s' f} \right) F' + \left( \frac{f''}{s'^2 f} + \frac{E - V}{s'^2} \right) F = 0 \ , \tag{8}$$

thus,

$$\frac{s''}{s'^2} + 2\frac{f'}{s'f} = \frac{\tilde{\tau}}{\sigma} \quad , \quad \frac{f''}{s'^2 f} + \frac{E-V}{s'^2} = \frac{\tilde{\sigma}}{\sigma^2} \quad . \tag{9}$$

Using the spirit of our earlier applications [13,14], the energy and potential terms in (9) are decomposed in two pieces, which will provide a clear understanding for the individual contributions of the $F$ and $f$ terms to the whole of the solutions, such that $E-V = (E_F + E_f) - (V_F + V_f)$. Therefore, the second equality in Eq. (9) is transformed to a couple of equation

$$\frac{f''}{f} = V_f - E_f \quad , \quad -\frac{\tilde{\sigma}}{\sigma^2} s'^2 = V_F - E_F \quad , \tag{10}$$

where $f$ can be expressed in an explicit form due to the first part in (9)

$$f(r) \approx (s')^{-1/2} \exp\left[\frac{1}{2}\int^{s(r)}(\tilde{\tau}/\sigma)ds\right] . \tag{11}$$

We note that Eqs. (10) and (11) are the significant piece of the present work, suggesting an improved formalism for the calculations performed within the N-U theory. In such a way that for a given polynomial $(F)$, the transformation function $(s)$ in (10) and subsequently $f$ in (11) are easily defined since, in principle, the corresponding $\sigma, \tilde{\sigma}$ and $\tilde{\tau}$ terms are well known. Consequently, by the elemantary calculations, right-hand sides of two equations in (10) reveal explicitly the forms of solvable potentials $(V_f + V_F)$ and their full energy spectrum $(E_f + E_F)$ related to $F$ interested. Finally, from (6), the corresponding wave function is readily obtained for the whole spectrum. This refined scenario proposed here neither involve a defect in the calculations, like the restriction $(\tau' \prec 0)$ in the usual N-U treatment, nor a tedious and cumbersome calculation process.

## 3. Application

For the illustration, we restrict ourselves to the orthogonal polynomials of mathematical physics since in this work we are interested only in bound state wave functions. In addition, for clarity, our treatment takes only the Jacobi and generalized Laguerre polynomials $(P_n^{(\alpha,\beta)}(s), L_n^{(\alpha)}(s))$ into account. The other orthogonal polynomials such as Gegenbauer, Chebyshev and Legendre can be obtained as special cases from $P_n^{(\alpha,\beta)}(s)$, thus the application of the procedure to these and others does not cause any problem.

## 3.1. Non-relativistic consideration

We first apply the model described above to the Jacobi polynomials. There is an intimate relationship between the Jacobi polynomials and the hypergeometric function [15]. Any wave function expressed in terms of Jacobi polynomials can also be expressed in terms of hypergeometric functions as well. Nevertheless, in some cases it is more convenient to use these polynomials, because a wide class of exactly solvable potentials can be found more easily if we take them as a starting point, as is the case here. At the same time, the aspect of the present consideration is more transparent in the case of orthogonal polynomials.

The consideration of the Jacobi polynomials will also convince the reader that the new formalism naturally removes the drawback inherited from the original theory, and make clear the inter-relation between Eqs (1) and (2) in view of the calculation process. It is reminded at this point that some researchers, for instance [3], used Eq. (2) instead of Eq. (1) unlike the others in their calculations. Our careful study, in connection with this, show that the consideration of particular polynomials discussed below reduces (1) to (2) because $\tilde{\sigma} = \Lambda \sigma$, where $\Lambda$ is a constant connected to the energy spectrum, and $\pi \to 0$, $\tilde{\tau} \to \tau$ and $F \to y$ for this choice. The reader is refered to Eqs. (1), (2), (8) and (9) for a deeper understanding of this point. In this case, it would be interesting to see the $\Lambda_n$ − value appearing in Eq. (3), which provides a testing ground for the reliability of the procedure underlined.

From the differential equation of the Jacobi polynomials related to Eq. (2), one sees that [15]

$$\sigma = 1 - s^2 \quad , \quad \tau = (\beta - \alpha) - (\alpha + \beta + 2)s \quad , \quad \Lambda_n = n(n + \alpha + \beta + 1) \quad , \tag{12}$$

which fulfil the requirement of the standard formalism, $\Lambda_n = -n\tau'(s) - \frac{n(n-1)}{2}\sigma''(s)$, since here $\tau' \prec 0$. From Eq. (10), $V_F - E_F = -\frac{\Lambda_n}{\sigma}s'^2$ because $\tilde{\sigma} = \Lambda\sigma$ in this polynomial choice. Since we have a constant $(E_F)$ on the left-hand side, there must be at least one term on the right-hand side, from which a constant arises. In the most general case this must be one of the terms containing the parameters $n$, $\alpha$ and $\beta$ of the Jacobi polynomials. Therefore, if we set

$$\frac{s'^2}{1-s^2} = C, \tag{13}$$

where $C = a^2$ is a positive constant, we can get different kinds of $s(r)$ functions from this simple differential equation, depending on the sign of $C$. A more detailed investigation regarding a similar search can be found in the famous work of Levai [12]. Chosing one of the five possible solutions of (13), $s(r) = \cos(ar)$, one arrives at

$$E_F = a^2 n(n + \alpha + \beta + 1) \quad , \quad V_F = 0 \quad , \tag{14}$$

and from (11), in which $\tilde{\tau} \to \tau$,

$$f = (-a)^{-1/2}(1-s)^{(2\alpha+1)/4}(1+s)^{(2\beta+1)/4} \quad , \tag{15}$$

leading to

$$\frac{f''}{f} = V_f - E_f \quad , \quad E_f = \frac{a^2}{4}(\alpha + \beta + 1)^2 \quad ,$$

$$V_f = \frac{a^2}{4}\left[(\alpha-\beta)^2 + (\alpha+\beta)^2 - 1\right]\cos ec^2(ar) + \frac{a^2}{2}(\alpha^2 - \beta^2)\cos ec(ar)\cot(ar) \quad . \tag{16}$$

Hence, the full energy spectrum and corresponding wave function for the total potential above, $V = V_F + V_f$, can be readily expressed by $E = E_F + E_f$ and $\Psi = fF$ where $F = P_n^{(\alpha,B)}$ being the generalized Jacobi polynomial. The results obtained are in agreement with those in [12].

Now, let us concentrate on a specific example which cannot be solved by the usual N-U method due to the sign of $\tau' \succ 0$ which takes to unacceptable energy values in the calculations. For the clarification of this point, we now consider another type of the Jacobi polynomial, $(1-s)^\alpha(1+s)^\beta P_n^{(\alpha,\beta)}$, for which

$$\sigma = 1 - s^2 \quad , \quad \tau = (\alpha - \beta) - (\alpha + \beta - 2)s \quad , \quad \Lambda_n = (n+1)(n + \alpha + \beta) \quad , \quad s = \cos(ar) \quad , \tag{17}$$

in this case one obtains,

$$E_F = a^2(n+1)(n + \alpha + \beta) \quad , \quad E_f = \frac{a^2}{4}(\alpha + \beta - 1)^2 \quad ,$$

$$V_f = \frac{a^2}{4}\left[(\alpha-\beta)^2 + (\alpha+\beta)^2 - 1\right]\cos ec^2(ar) + \frac{a^2}{2}(\alpha^2 - \beta^2)\cos ec(ar)\cot(ar) \quad , \quad V_F = 0. \tag{18}$$

Though the resulting energy expressions in the both example seem different, they are in fact identical. This is not surprising indeed since the corresponding potentials $(V = V_F + V_f)$ and wave functions are the same. Nevertheless, the apperance of $E_F$ in (14) is more suitable than that of in (18) due to the properties of orthogonal polynomials and hypergeometric functions in case $n \to 0$. This consideration once makes clear the superiority of the new procedure when compared to the usual treatment in the N-U model which fails in solving the present example, because $\Lambda_n = -n\tau'(s) - \frac{n(n-1)}{2}\sigma''(s)$ is not valid in this case.

Applying the same procedure to the generalized Laguerre polynomials related to confluent hypergeometric functions

$$\sigma = s \quad , \quad \tau = s+1 \quad , \quad \tilde{\sigma} = \left(n+\frac{\alpha}{2}+1\right)s - \frac{\alpha^2}{4} \quad , \tag{19}$$

one obtains, from (10),

$$s = \frac{a^2}{4}r^2 \quad \Rightarrow \quad V_F = \frac{\alpha^2}{r^2} \quad , \quad E_F = a^2\left(n+\frac{\alpha}{2}+1\right) \quad , \tag{20}$$

and deduces that

$$f = \left(\frac{a^2 r}{2}\right)^{-1/2} e^{s/2} s^{1/2} \quad \Rightarrow \quad V_f = \frac{a^4}{16}r^2 - \frac{1}{4r^2} \quad , \quad E_f = -\frac{a^2}{2} \quad , \tag{21}$$

leading to the well known harmonic oscillator potential $(V_F + V_f)$ in three-dimension, where $\alpha = \ell + \frac{1}{2}$ and $a^2 = 2w$. The corresponding unnormalized wave function is $\Psi = fF = s^{(2\alpha+1)/4} e^{-s/2} L_n^\alpha$. In the present example $V_F \neq 0$, unlike the former case, because now the $\tilde{\sigma}$ term has a different structure.

Similar considerations can be applied to other special functions satisfying a homogeneous linear second-order differential equation, however such applications, which are not presented here, do not reveal anything new in view of the present discussion.

**3.2. Relativistic consideration**

This section involves an attempt to extend the same scenario to the relativistic region, for which the $s$-wave Klein-Gordon (K-G) equation leading to bound states is considered. In the presence of vector and scalar potentials the (1+1)-dimensional time-independent K-G equation for a spinless particle of rest mass $m$ reads $(\hbar = c = 1)$

$$-\psi'' + (m+V_s)^2 \psi = (\varepsilon - V_v)^2 \psi \quad , \tag{22}$$

in which $\varepsilon$ is the relativistic energy of the particle, together with $V_v(r)$ and $V_s(r)$ being the vector and scalar potentials respectively. Gaining confidence from the work in [16], we suggest that the full relativistic wave function in (22) can be expressed, as in the case of the

Schrödinger equation above, by $\psi(r) = F(s)f(r)$ where $F$ now denotes the behaviour of the wave function in the non-relativistic region while $f$ represents the modification function due to the relativistic effects. This transformation reproduces similar results to those in section 2 with some slight differences such as the one that $E$ and $V$ terms in Eqs. (7-9) now turns out to be

$$E \to \varepsilon^2 - m^2 \quad , \quad V \to 2(mV_s + \varepsilon V_v) + (V_s^2 - V_v^2) \, , \tag{23}$$

leading to

$$-\frac{\tilde{\sigma}}{\sigma^2}s'^2 = 2(mV_s + \varepsilon V_v) - \varepsilon_F \quad , \quad \frac{f''}{f} = (V_s^2 - V_v^2) - \varepsilon_f \, . \tag{24}$$

Here, $\varepsilon_F$ and $\varepsilon_f$ represent, respectiveley, the energy in the non-relativistic limit and energy correction due to the relativistic consideration. For the calculation of $f$ in (24), one should use Eq. (11). The first part in [24] is the exact appearence of the K-G equation in the non-relativistic domain [16], which is expressed explicitly in terms of orthogonal polynomials, whereas the second equality generates the relativistic modifications via a properly constructed $f-$function. If the scaler and vector potentials are equal to each other $(V_s = \pm V_v)$, the relativistic corrections die away and $\psi \to F$ since $f \to$ constant.

Let us illustrate this discussion with a simple example. Before proceeding, we should remark that the scheme introduced here is distinct from the Schrödinger applications presented in the previos sections. As the potentials in (24) are in principle known through the K-G procedure, unlike the Schrödinger case, for which one should define in this case a proper $s-$function for the transformation to satisfy left-hand sides of the equations in (24) to extract the related energy values and wave functions. To achive our goal, we focus on the problem of a particle subject to an inversely linear potential in one spatial dimension, which has received considerable attention in the literature, for a recent review see [17]. In this work, the mixed vector-scalar inversely linear potentials are in the form of

$$V_v = -\frac{A}{|r|} \quad , \quad V_s = -\frac{B}{|r|} \, , \tag{25}$$

in which the coupling constants, $A$ and $B$ are dimensionless real parameters. Choosing the most appropriate Hermite polynomial $\left(e^{-s^2/2}H_n(s)\right)$

$$\sigma = 1 \quad , \quad \tilde{\tau} = 0 \quad , \quad \tilde{\sigma} = 2n + 1 - s^2 \, , \tag{26}$$

it is not hard to see that $s'^2 s^2 = a^2$ leading to $s = \sqrt{2a|r|}$. For clarity, we consider only the positive values in the square root to deal with the real potentials, then

$$V_F = -\frac{a(2n+1)}{2|r|} \quad , \quad E_F = -a^2 \quad , \quad V_f = -\frac{3}{16r^2} \quad , \quad E_f = 0 \quad , \tag{27}$$

for which the corresponding wave function is $\psi = Ff = s^{1/2} e^{-s^2/2} H_n$. These results agree with those in Ref. [12,17]. The Kratzer-like potential with singularity given by $-1/r^2$, which is comprehensively discussed in [17] with the consideration of bound and unbound state possibilities, contains $n-$ term in its first piece. To end up with reasonable results, one has to shift the $n-$dependence to the energy value and rid the remaining terms of $n$. This can be carried by a transformation of the parameters such that $C = a(2n+1)/2$, which determine the $n-$dependence of the relativistic energy spectrum,

$$\varepsilon^2 - m^2 = \varepsilon_F + \varepsilon_f = -a^2 = -\frac{4C^2}{(2n+1)^2} = -\frac{16(mB + \varepsilon A)^2}{(2n+1)^2} \quad , \tag{28}$$

where $A = \pm\sqrt{B^2 + 3/16}$ that supports the related analysis in [17].

## 4. Concluding Remarks

Here, we have investigated a simple method of finding solvable potentials in non-relativistic quantum mechanics. The formalism systematically recovers known results in a natural unified way and allows one to extend certain results known in particular cases. This method can be used to transform the Schrödinger equation into a linear homogeneous second-order differential equation with known special functions as solutions, which refines the calculations performed within the frame of the Nikiforov-Uvarov method. The procedure also has been extended successfully for the Klein-Gordon equation to express explicitly the solutions at the non-relativistic limit and the corrections due to relativistic consideration. A straightforward generalization would be the application of the scheme to the Dirac equation. Beyond its intrinsic importance as a new solution for a fundamental equation in physics, we expect that the present simple method would find a widespread application in the study of different quantum mechanical and nuclear scattering systems. Along this line the works are in progress.

# REFERENCES

**[1]** Nikiforov A F and Uvarov V B 1988 *Special Functions of Mathematical Physics* (Basle: Birkhauser)

**[2]** Eğrifes H, Demirhan D and Büyükkılıç F 1999 *Phys. Scr.* **59** 90; **60**, 195

**[3]** Cotfas N 2002 *J. Phys. A: Math. Gen.* **35** 9355; 2004 *Cent. Eur. J. Phys* **2** 456

**[4]** Yeşiltaş Ö, Şimşek M, Sever R and Tezcan C 2003 *Phys. Scr.* **67** 472

**[5]** Şimşek M and Eğrifes H 2004 *J. Phys. A: Math. Gen.* **37** 4379

**[6]** Aktaş M and Sever R 2004 *J. Mol. Struc. (Theochem)* **710** 223

**[7]** Berkdemir A, Berkdemir C and Sever R 2004 *Phys. Rev*. C **72** 27001

**[8]** Yaşuk F, Berkdemir C and Berkdemir A 2005 *J. Phys. A: Math. Gen.* **38** 6579

**[9]** Yaşuk F, Berkdemir C, Berkdemir A and Önem C 2005 *Phys. Scr.* **71** 340–3

**[10]** Berkdemir C, Berkdemir A and Sever R 2006 *nucl-th/0501030 v4*

**[11]** Berkdemir C 2006 *Nucl. Phys. A* **770** 32

**[12]** Levai G 1989 *J. Phys. A: Math. Gen.* **22** 689

**[13]** Gönül B, Köksal K 2006 *quant-ph/0512216*

**[14]** Gönül B, Köksal K and Bakır E 2006 *Phys. Scr*. **73** 279

**[15]** Abramowitz M and Stegun I A 1970 *Handbook of Mathematical Functions* (New York: Dover)

**[16]** Gönül B 2006 *quant-ph/0603181*

**[17]** S. de Castro A 2005 *Phys. Lett. A* **338** 81